# Polarization conversion using hybrid near-zero and high-index metamaterials

Ben Johns[1,2]\*, Hannes Kempf[1], Lakshmi Das[1], Aitor de Andrés[1], and Nicolò Maccaferri[1,2]\*

[1]*Department of Physics, Umeå University, 90187 Umeå, Sweden*

[2]*Umeå Centre for Microbial Research, 90187, Umeå, Sweden*

ben.johns@umu.se, nicolo.maccaferri@umu.se

Near-zero-refractive index materials display unique optical properties such as perfect transmission through distorted waveguides, cloaking, and inhibited diffraction. Compared to conventional media, they can fundamentally behave differently to light impinging from air, owing to the phenomenon of total external reflection. This makes them attractive for evanescent wave phenomena without prism coupling or sub-wavelength air gaps. Here, we introduce a strategy for linear-to-circular polarization conversion based on this effect. Practically realizable device designs are presented using hybrid structures of near-zero index and high index materials in a simple planar geometry. Our results predict polarization conversion in both reflection and transmission configurations based on naturally available epsilon-near-zero materials and illustrate the role of loss in limiting performance. Although the operation wavelength is limited to the epsilon-near-zero region, the concept can be implemented from the terahertz to optical domains using tunable epsilon-near-zero materials.

Manipulating the polarization of light is of utmost importance in fast-developing fields such as holography,[1] biomedical imaging,[2] and polarization multiplexing.[3] Versatile control over polarization requires approaches for strengthening light-matter interactions, especially at sub-wavelength length scales. Recently, near-zero-refractive index (NZI) media have emerged as a platform for unique and enhanced light-matter interactions,[4] with potential applications such as enhanced non-linear processes,[5–9] waveguide super-coupling,[10] ultrafast switching,[11–13] sensing,[14] non-reciprocal light propagation[15] and directional emission.[16] Moreover, NZI structures have been proposed as waveguides and other components in integrated photonics.[17] However, NZI-based approaches for polarization manipulation typically rely on anisotropic metamaterials[18,19] or by integrating polarization-selective metasurface elements with NZI structures,[20] which are typically fabrication-demanding. Moreover, a unifying design concept that is applicable irrespective of the operating frequency, for example from the terahertz to the optical domain, is missing.

Here, we present a strategy for polarization conversion based on naturally available NZI – high index layered materials. Our approach exploits the very low allowed wave numbers for light in an NZI medium[16] given by $k_{NZI} = n_{NZI} k_0$, where $n$ is the refractive index, $k_0 = \omega/c$ is the wave number of light in free space, $\omega$ is the angular frequency and $c$ is the speed of light. For light falling on an air-NZI interface at an angle of incidence (AOI) $\theta$, the in-plane wave vector component $k_{||} = k_0 \sin\theta$ becomes greater than $k_{NZI}$ above the critical angle $\theta_c = \text{asin}\, n_{NZI}$.[21] Since $\theta_c$ can be small, light becomes evanescent within the NZI medium at experimentally accessible angles;[22,23] for example, $\theta_c = 30°$ when $n_{NZI}$ is 0.5. This unique situation where air acts as the *high-index* medium allows evanescent wave phenomena to occur in planar optical stacks without bulky light in-coupling elements or precisely controlled sub-wavelength air gaps, with the potential for compact and integrated device functionality.

To introduce the physical basis of our concept for polarization conversion, we first discuss the simpler (ideal) case of lossless NZI materials. Above the critical angle, light can either be totally reflected (total external reflection or TER, schematic in **Figure 1a**),[21] or partially transmitted if the NZI medium is thin compared to the wavelength (optical tunneling or OT, schematic in **Figure 1b**).[24] While undergoing either TER or OT, orthogonally polarized components of light (*s* and *p*) experience different, controllable phase shifts ($\delta_s$ and $\delta_p$). The relative phase shift between the two components is given by $\Delta = \delta_s - \delta_p$. When $\Delta$ = 0°, an incident linearly polarized wave remains as such and when $\Delta$ = 90°, it is converted to circularly polarized light (for other values of $\Delta$, the wave is elliptically polarized). It was previously shown that achieving evanescent wave-based linear-to-circular polarization conversion in TER (reflection mode, $\Delta_r$ = 90°) or OT (transmission mode, $\Delta_t$ = 90°) requires a refractive index contrast $n_{12} = n_2/n_1 < \sqrt{2} - 1$ = 0.414. [24–27] Here, $n_1$ is the high-index medium from which light is incident and $n_2$ is the low-index medium (see schematics in **Figure 1a,b**). One of the earliest devices for linear-to-circular polarization conversion was the Fresnel rhomb, which made use of this principle (**Supplementary note 1**). Current approaches where air is always the low-index medium ($n_2 = 1 \Rightarrow n_1 > 2.414$) face the challenge of coupling the light into a high-index incident medium (typically glass or other high-index dielectrics) using prisms.[28] Additionally, in the OT configuration, the use of air gaps as the low-index medium requires precisely-controlled sub-micron spacing between two prisms.[29] In our approach, by employing NZI materials the index contrast condition ($n_{12} < 0.414$) can be satisfied with *air* as the high-index medium if $n_2 = n_{NZI} < 0.414$. In addition, w e use dielectric layers as the high-index medium to further minimize $n_{12}$ (= $n_{NZI}/n_d$), where $n_d > 1$ is the dielectric refractive index. Such a multilayer design incorporating low-index and high-index media can then potentially allow polarization conversion when light is incident directly from air.

To illustrate this further, the maximum value of $\Delta$ on reflection, $\Delta_{r,max}$, for light incident above $\theta_c$ at an interface is given by [24]

$$\tan \frac{\Delta_{r,max}}{2} = \frac{1-n_{12}^2}{2n_{12}} \qquad (1)$$

Note that following the $e^{-i\omega t}$ convention, the phase advance after reflection or transmission ($\delta_p, \delta_s$) is taken as the negative of the phase of the corresponding Fresnel reflection or transmission coefficient.[30]

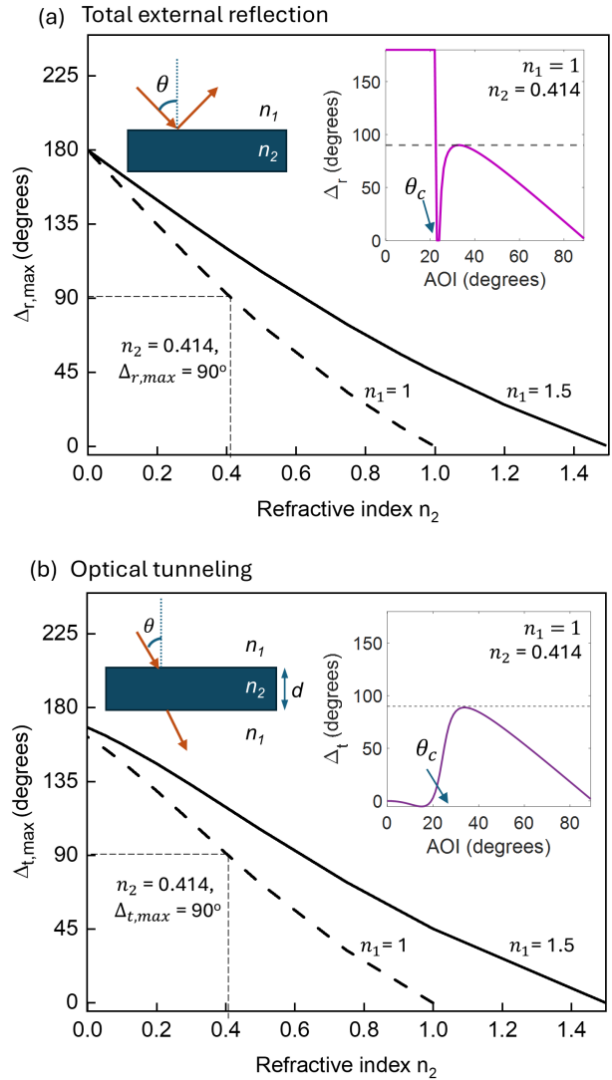

Fig.1. (a) The maximum value of Δ on reflection, $\Delta_{r,max}$, plotted as a function of the refractive index $n_2$ (see schematic inset of reflection mode configuration) for $n_1 = 1$ (dashed) and $n_1 = 1.5$ (solid curve). The plot inset shows $\Delta_r$ for $n_1 = 1$ and $n_2 = 0.414$ as a function of AOI. (b) $\Delta_{t,max}$, the maximum value of Δ in transmission mode configuration (see schematic) plotted as a function of $n_2$ for $d/\lambda = 1$. The plot inset shows $\Delta_t$ for $n_1 = 1$ and $n_2 = 0.414$ as a function of AOI.

**Figure 1a** plots $\Delta_{r,max}$ as a function of $n_2$ for $n_1 = 1$ and 1.5. The plot shows that $\Delta_{r,max} > 90°$ can be achieved for incidence from air with a single reflection if $n_2 < 0.414$ (dashed curve). The inset in **Figure 1a** shows the variation in $\Delta_r$ with AOI for $n_2 = 0.414$. For angles below $\theta_c$ (here 24.45°), $\Delta_r$ is either 0 or

180°. Above $\theta_c$, the dependence of $\delta_s$ and $\delta_p$ on AOI results in a continuous variation of $\Delta_r$ reaching its maximum value (i.e., $\Delta_{r,max}$) of 90°. Notably, when $n_1 = 1.5$ (solid curve in **Figure 1a**), the index contrast is further improved. This relaxes the required NZI index to achieve $\Delta_{r,max}$ of 90° from 0.414 to 0.621, which will play a significant role later when we consider real materials. Analogously, the case of optical tunneling is shown in **Figure 1b**. The variation in $\Delta_{t,max}$ (the maximum achievable $\Delta$ in transmittance) is plotted in **Figure 1b** a function of $n_2$ with $n_1 = 1$ and 1.5. We see that $\Delta = 90°$ is achievable for $n_2 \lesssim 0.414$ when $n_1 = 1$ (dashed curve). For $n_1 = 1.5$, the required NZI index is relaxed as expected (solid curve). Here, $\Delta_{t,max}$ additionally depends on the thickness of the tunneling layer, $d$, which is equal to the wavelength $\lambda$ in **Figure 1b** (Supplementary Note 2).[29] The inset in **Figure 1b** shows the dependence of $\Delta_t$ on AOI for $n_2 = 0.414$, where it reaches $\Delta_{t,max} \approx 90°$ above the critical angle. This demonstrates that $\Delta_r$ and $\Delta_t$ values large enough for linear to circular polarization conversion are achievable based on TER and OT in hybrid low-index and high-index media.

Having demonstrated the physical principle using lossless NZI media, we next explore possible practical realizations of this concept. Epsilon-near-zero (ENZ) materials, which show near-zero refractive indices around their ENZ wavelengths if material losses (represented by the imaginary part of the refractive index) are low enough,[6] are potential candidates to implement this concept. Two factors need to be considered when extending the idea to real materials: dispersion and loss. Material dispersion limits the operation range of wavelengths for a given ENZ material to the vicinity of its ENZ wavelength $\lambda_{ENZ}$, where a near-zero refractive index may be available.[22,31] Moreover, losses may limit the achievable relative phase advances $\Delta_t$ or $\Delta_r$ when compared to the lossless case. It should be noted that in lossy ENZ media, the reflection is no longer 'total' above the critical angle,[22] and we use the term attenuated total external reflection (ATER) to refer to the phenomenon in the presence of loss. We chose silicon carbide (SiC) and n-doped cadmium oxide (CdO), which are low-loss ENZ materials, for our numerical calculations. Extreme field confinement [32] and ultrafast nonlinear response [33] has been reported in the ENZ regimes of these materials. The real part of the permittivity ($\epsilon'$) of SiC approaches zero at its longitudinal optical (LO) phonon wavelength [32,34,35] at 10.32 μm (**Figure 2a**). At longer wavelengths, SiC is metallic due to its Reststrahlen band.[32] Below the LO phonon wavelength in its dielectric regime, SiC behaves as a mid-IR ENZ material.[32] Notably, the refractive index of SiC lies below 1 for wavelengths between $\lambda$ = 10.03 µm and 10.32 µm, with low associated losses, allowing the possibility of ATER and OT. Similarly, n-doped CdO is a low-loss ENZ material with a tunable $\lambda_{ENZ}$ in the near- and mid-IR ranges.[33] Here, we take a near-IR value of $\lambda_{ENZ}$ = 1.9 µm, with a corresponding $n$ < 1 window between $\lambda$ = 1.71 µm and 1.9 µm (**Figure 2b**).[36,37]

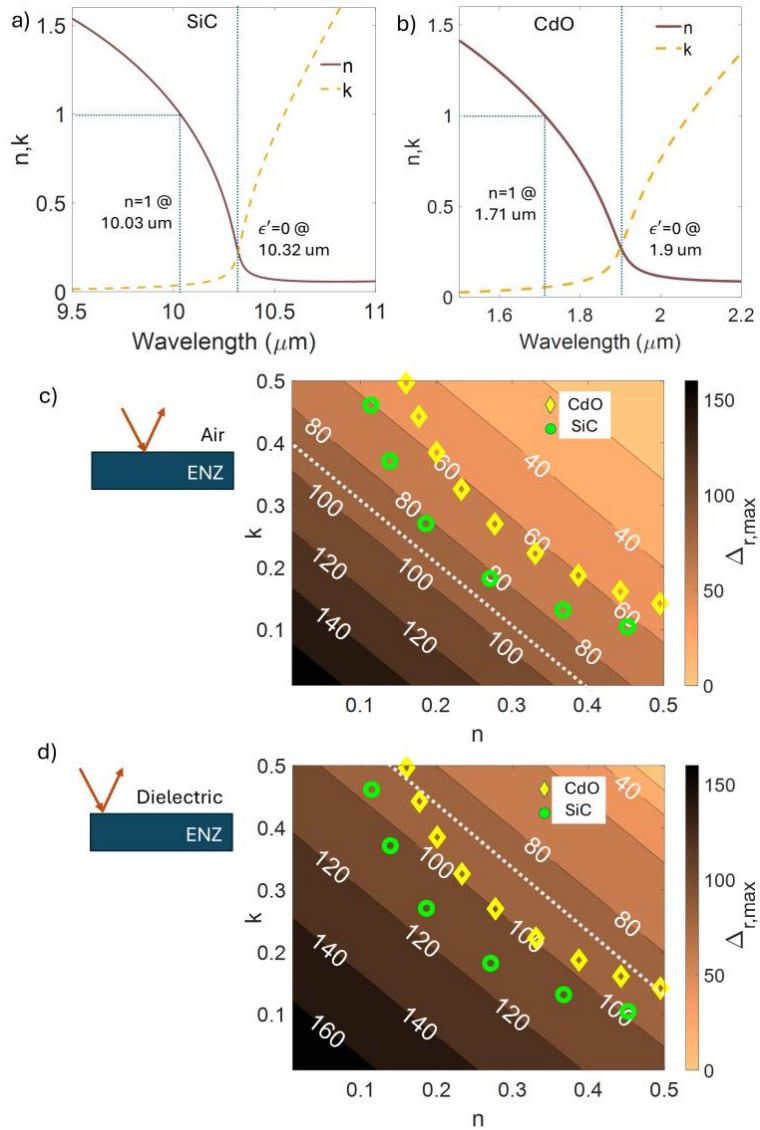

Fig. 2. Refractive index of (a) silicon carbide and (b) n-doped cadmium oxide around their ENZ wavelengths. Vertical lines in (a) and (b) demarcate the low-loss ENZ regime between the ENZ wavelength and the $n = 1$ wavelength. (c, d) Contour plots of $\Delta_{r,max}$ (in degrees) as a function of $n$ and $k$ of the ENZ medium for the ATER configuration with incidence from (c) air and (d) dielectric with $n_d = 1.5$ (see schematic in inset). The $\Delta_{r,max}$ values for CdO and SiC in their respective ENZ regimes are indicated by symbols, and the blue diagonal line indicates $\Delta_{r,max} = 90°$.

We first consider the case of ATER for light incident from air onto a semi-infinite ENZ material (**Figure 2c**). To include the effect of loss, we calculated the achievable $\Delta_{r,max}$ as a function of the ENZ $n$ and $k$ (extinction coefficient) for this configuration. From **Figure 2c**, we observe that lower values of $n$ and $k$ improve the $\Delta_{r,max}$ values. The achievable $\Delta_{r,max}$ values for SiC and CdO in their ENZ regimes are plotted for comparison, showing that $\Delta_{r,max}$ = 90° (denoted by the dotted diagonal line) cannot be achieved using these materials for light falling directly from air. In contrast, considering a dielectric of $n_d$ = 1.5 as the ambient incident medium (we consider a semi-infinite dielectric now, a more practical structure based on a dielectric coating is shown later), the calculated $\Delta_{r,max}$ increases above 90° (**Figure 2d**) for both CdO and SiC. Similarly, we also investigated ENZ materials in the optical tunneling configuration (**Figure 3**). Here, we present results of our calculations for SiC (similar results for CdO are reported in **Supplementary Note 3**). **Figure 3a** shows $\Delta_{t,max}$ for light incident from air on a SiC layer on silicon substrate as a function of wavelength and SiC thickness. The contour map reveals interesting trends in $\Delta_{t,max}$ when using real ENZ materials. On the short wavelength side, $\Delta_{t,max}$ remains small where $n_{SiC}$ > 1 and has appreciable values only once $n_{SiC}$ falls below 1, as expected. Further, a larger value of $\Delta_{t,max}$ is possible closer to $\lambda_{ENZ}$ as the refractive index of SiC decreases providing an improved index contrast with the incident medium. The value of $\Delta_{t,max}$ is also improved by using silicon substrate, which has a high refractive index of 3.4.[38] For wavelengths at or longer than $\lambda_{ENZ}$, $\Delta_{t,max}$ again becomes small when high losses strongly suppress the phase shifts. These calculations verify that the large $\Delta_{t,max}$ predicted for lossless NZI media can be also realized near the ENZ wavelength in real ENZ materials. We also observe a gradual increase in $\Delta_{t,max}$ with thickness of the SiC layer. However, the maximum $\Delta_{t,max}$ obtainable lies below 90° in **Figure 3a**; using thicker SiC layers becomes impractical as it strongly reduces the intensity of light tunneling through it. In contrast, by using a high index ambient medium having $n_d = 1.5$, we observe a marked improvement in the achievable $\Delta_{t,max}$ (**Figure 3b**). Indeed, $\Delta_{t,max}$ of 90° is readily possible within the ENZ regime for SiC thicknesses above 3.5 μm or $d/\lambda = 0.33$.

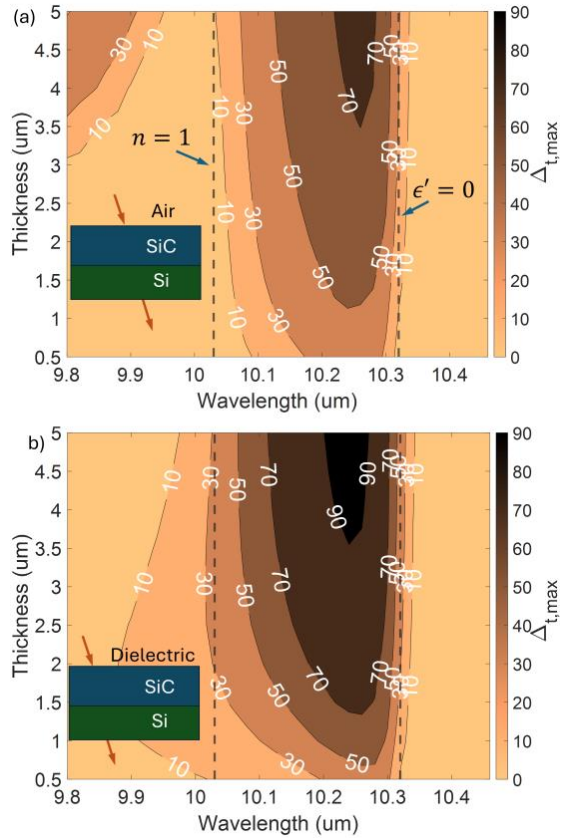

Fig. 3. (a,b) Contour plots of $\Delta_{t,max}$ (in degrees) as a function of wavelength and ENZ thickness for SiC on a silicon substrate with ambient medium (a) air and (b) $n_d = 1.5$. Insets show the schematic of the corresponding configuration. The low-loss ENZ regime between the ENZ wavelength and the $n = 1$ wavelength are demarcated by the vertical dashed lines.

The above results (**Figure 2** and **Figure 3**) suggest that improving the index contrast may allow linear-to-circular polarization conversion by ATER and OT in real, lossy ENZ materials. However, to maintain free space operation, the ambient medium should be air. To address this, one strategy is to apply a high-index dielectric coating to the ENZ layer (see schematic in **Figure 4a,b**). For example, $n_d \approx 1.5$ is obtainable by thin film deposition of SiO$_2$ or solution-processible IR transparent polymers such as polyethylene. We note that regardless of the presence of the dielectric coating, light falling from air onto the layered structure at an angle above $\theta_c$ becomes evanescent in the ENZ layer. To see this, the condition for light to be evanescent is $n_{ENZ} k_0 < k_{||}$, where $n_{ENZ}$ is the real part of the ENZ medium refractive index. Since $k_{||}$ remains the same throughout the multilayer stack, this condition becomes $n_{ENZ} k_0 < k_0 \sin \theta$, where $\theta$ is the AOI from air, which is identical to the critical angle condition without an intervening layer. To demonstrate the practical feasibility of such an optical device, we present results for SiC (similar results for CdO are reported in **Supplementary Note 4**). In the ATER configuration (**Figure 4a**), we assume semi-infinite SiC thickness and $n_d = 1.5$ at a wavelength of 10.23 µm. In the absence of the dielectric ($t_d = 0$ nm), $\Delta_r < 90°$ as noted earlier. Note that although $\Delta_r = 90°$ is satisfied at lower AOI within the shaded region ($\theta < \theta_c$), the reflectance is extremely low here as it lies near the pseudo-Brewster angle and thus is not suitable for our application. Adding a thin dielectric layer increases the refractive index contrast at the dielectric-SiC interface, thus improving $\Delta_r$. The value of $\Delta_r$ increases gradually with $t_d$ at all angles and its maximum value $\Delta_{r,max}$ crosses 90° for $t_d \approx 400$ nm and above. The corresponding results in the OT configuration with Si substrate are reported in **Figure 4b**. In this case as well, $\Delta_t$ is below 90° when $t_d = 0$ nm. On adding a dielectric layer ($n_d = 1.5$), $\Delta_t$ increases as the index contrast improves and $\Delta_{t,max}$ crosses 90° for $t_d = 900$ nm and above. Note that the specific dielectric thickness required for $\Delta_{t,max}$ to reach 90° depends on the thickness of the ENZ layer used ($d = 3$ µm here). Finally, the main limitations of this design are limited bandwidth due to dispersion and limited efficiency due to ENZ losses (see **Supplementary Note 5**).

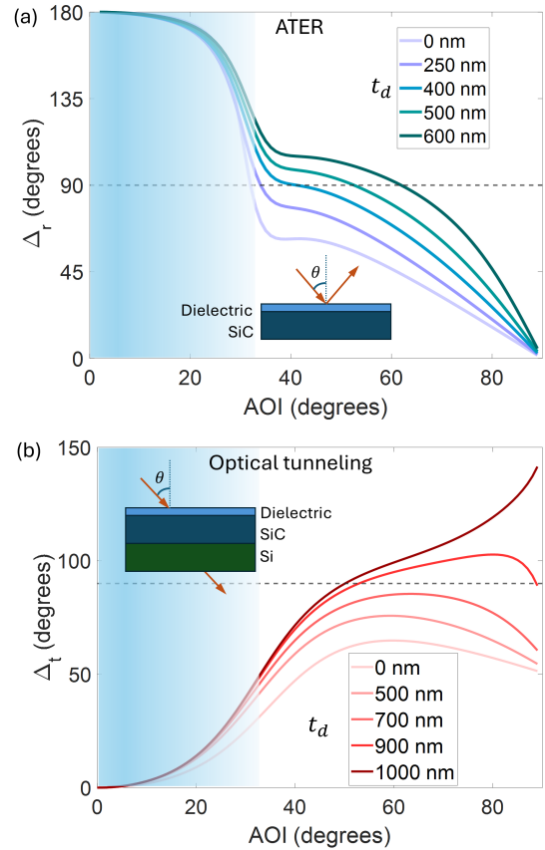

Fig. 4. Plot of (a) $Δ_r$ vs AOI in the ATER configuration and (b) $Δ_t$ vs AOI in the OT configuration with ENZ thickness $d = 3$ μm. The dielectric thickness $t_d$ is varied as indicated in the legend. In (a) and (b), $n_d = 1.5$ and $λ = 10.23$ μm. The shading indicates the region where $θ < θ_c$.

In summary, we have presented a novel strategy to achieve polarization conversion using a hybrid near-zero index – high index configuration based on evanescent wave generation. No coupling optics or sub-micron gaps are required in this design, which can work in both reflection and transmission configurations. Our concept makes use of isotropic, continuous ENZ materials and is thus different from previous ENZ-based polarization manipulation schemes based on anisotropic metamaterials.[18,19] Although we have demonstrated two examples (CdO in the near-IR and SiC in the mid-IR), the operating wavelength can be controlled using tunable ENZ materials available across the visible, infrared and terahertz wavelengths.[31,36,39–41] As a result, our strategy represents an example of how the unique properties of NZI materials can be utilized in flat-optics device design.

**Funding.** Swedish Research Council (Grant No. 2021-05784), Knut and Alice Wallenberg Foundation through the Wallenberg Academy Fellows Programme (Grant No. 2023.0089), European Research Council (ERC Starting Grant No. 101116253 'MagneticTWIST'), UCMR "Excellence by Choice" programme funded by Kempestiftelserna (Grant No. JCK-2130.3), European Union's Horizon 2020 Research and Innovation Programme under the Marie Skłodowska-Curie Actions (Grant No. 101147248).

**Disclosures.** The authors have no conflicts to disclose.

**Data Availability.** Data underlying the results presented in this paper are not publicly available at this time but may be obtained from the corresponding authors upon reasonable request.

**Supplemental Document.** See Supplementary Material for supporting content.

# Polarization conversion using hybrid near-zero and high-index metamaterials

BEN JOHNS[1,2]*, HANNES KEMPF[1], LAKSHMI DAS[1], AITOR DE ANDRÉS[1], AND NICOLÒ MACCAFERRI[1,2]*

[1]Department of Physics, Umeå University, 90187 Umeå, Sweden

[2]Umeå Centre for Microbial Research, 90187, Umeå, Sweden

ben.johns@umu.se, nicolo.maccaferri@umu.se

## Content

Supplementary Note 1: Fresnel's rhomb

Supplementary Note 2: Dependence of $\Delta_{t,max}$ on refractive index and thickness of NZI layer

Supplementary Note 3: Optical tunneling configuration – n-doped CdO in the near-IR range

Supplementary Note 4: Thin dielectric layer – n-doped CdO in the near-IR range

Supplementary Note 5: Note on efficiency and limitations of the design

## Supplementary Note 1: Fresnel's rhomb

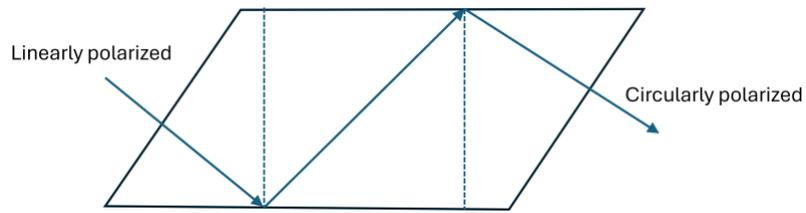

**Figure S1:** Schematic of linear to polarization conversion using two reflections in a Fresnel's rhomb.

For completeness, the working principle of a Fresnel's rhomb is included here (**Figure S1**). Linearly polarized light with electric field at 45° to the plane of incidence (i.e., the azimuth angle = 45°) is used to ensure that equal intensities are present in the *s* and *p* components. Due to total internal reflection from the sides of the prism, the reflected intensities will also be the same. To obtain $\Delta_{r,max}$ = 90°, an index contrast of $n_{12}$ = 0.414 is required (Eq. 1 in the main text). With air as the low-index medium ($n_2$ = 1), this would require a medium with $n_1$ = 2.414 which is difficult to achieve with the optical dielectrics such as glasses ($n \approx 1.5$). With glass, $\Delta_{r,max}$ is roughly 45°, therefore one can use two total reflections at the glass-air interface which each contribute $\Delta_r$ = 45° to accumulate the overall 90° phase shift between *s* and *p* polarizations [1].

## Supplementary Note 2: Dependence of $\Delta_{t,max}$ on refractive index and thickness of NZI layer

Figure 1b in the main text shows the variation in $\Delta_{t,max}$ with the refractive index contrast for a thickness to wavelength ratio $d/\lambda = 1$. It was then identified that a minimum index contrast of 0.414 is required to realize $\Delta_{t,max} \geq 90°$. Here, we show that this holds for any thickness. **Figure S2** shows $\Delta_{t,max}$ plotted against thickness and refractive index $n_2$ for a lossless thin layer surrounded by air (as considered in Figure 1b in the main text). At low thickness, the index contrast required for $\Delta_{t,max} = 90°$ is very small (around 0.1). This improves with thickness and reaches a limiting value of 0.414, verifying that an index contrast of 0.414 or better is required irrespective of NZI thickness.

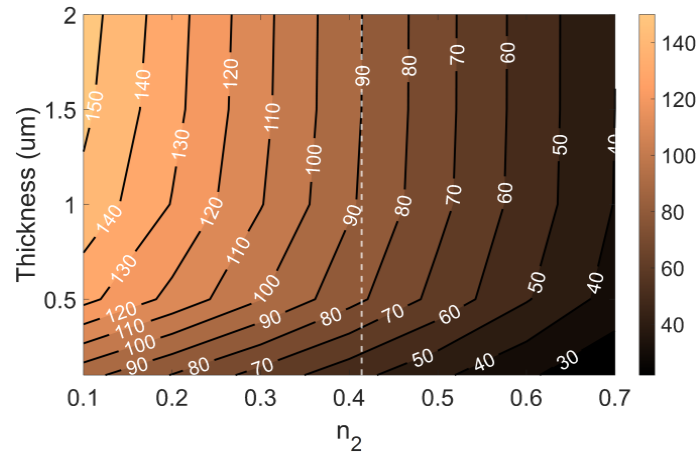

**Figure S2**: Contour plot of $\Delta_{t,max}$ as a function of NZI thickness and refractive index ($n_2$) for $n_1 = 1$ and $\lambda = 1$ $\mu$m. The dashed vertical line denotes $n_2 = 0.414$.

**Supplementary Note 3: Optical tunneling configuration – n-doped CdO in the near-IR range**

Similar to the results presented for SiC in the optical tunneling configuration (Figure 3 of main text), we also explored a near-infrared configuration using n-doped CdO as the ENZ material. **Figure S3** shows the contour maps of $\Delta_{t,max}$ for CdO on Si substrate. The contour is plotted as a function of CdO thickness and wavelength. We compared air and a dielectric medium with $n_d$ = 1.8 as the ambient dielectric. As in the case of SiC, the contour plots verify that the required phase shifts are available in the low-loss ENZ regime (denoted by the dashed vertical lines) only when using a dielectric as the incident medium. Notably, we found that a larger $n_d$ (1.8) is necessary here in comparison to the value of 1.5 used in case of SiC. This is attributed to the higher loss in CdO in its ENZ regime (see similar effect in reflection for SiC and CdO co-plotted in Figure 2 in the main text), which requires a larger index contrast to achieve the necessary phases.

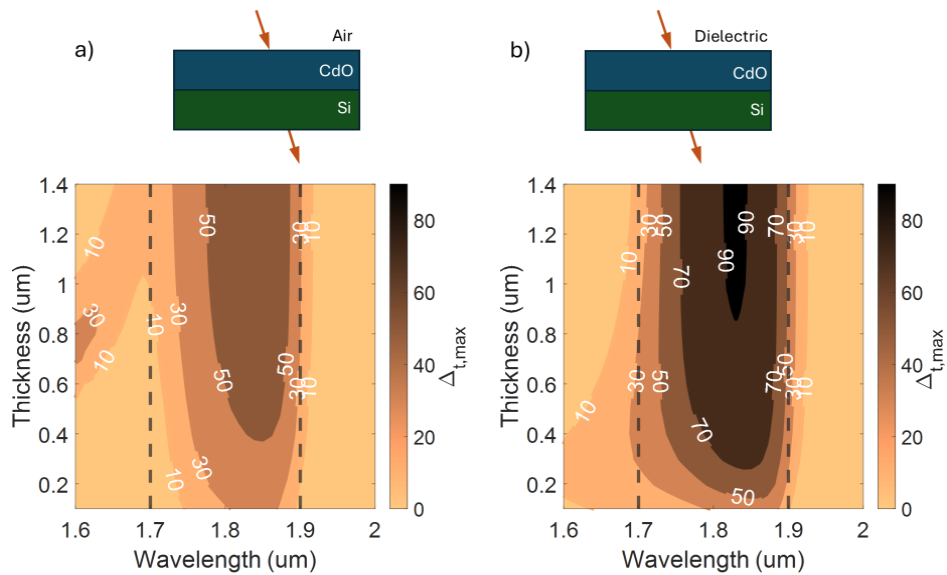

**Figure S3**: $\Delta_{t,max}$ for the optical tunnelling configuration with light incident from on the CdO layer on Si substrate from (a) air, (b) dielectric with $n_d$ = 1.8.

## Supplementary Note 4: Thin dielectric layer – n-doped CdO in the near-IR range

Similar to the results presented for SiC with a thin dielectric layer (Figure 4 of main text), here we report results for the analogous near-infrared device using n-doped CdO. **Figure S4a** plots $\Delta_t$ in the optical tunneling configuration with the thickness of n-doped CdO fixed at 500 nm. The thickness of the dielectric layer ($n_d$ = 1.8) is varied from $t_d$ = 0 nm to 200 nm, resulting in $\Delta_t$ increasing to values above 90º for $t_d$ above 100 nm. **Figure S4b** plots $\Delta_r$ in the ATER configuration. Here again, $\Delta_r$ lies below 90º when $t_d$ = 0 nm. On gradually increasing $t_d$ above 50 nm, the required larger phase shifts are obtained. These results for CdO in the near-IR, along with that of SiC in the mid-IR, demonstrate the generality of our proposed scheme.

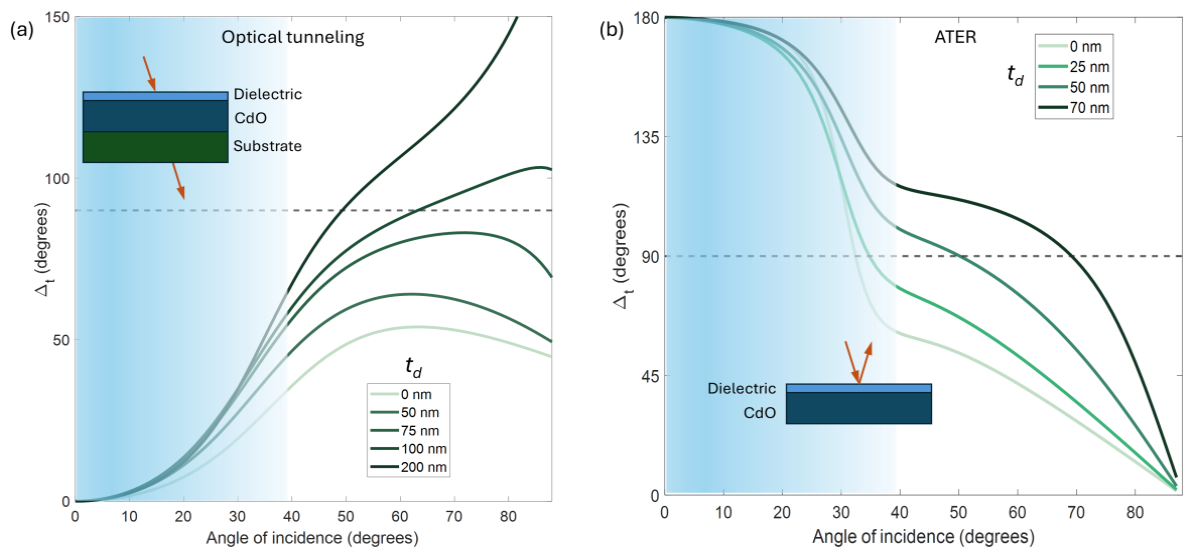

**Figure S4**. Plot of (a) $\Delta_t$ against AOI in the optical tunneling configuration with $t_{ENZ}$ = 500 nm and (b) $\Delta_r$ against AOI in the ATER configuration. The dielectric thickness $t_d$ is varied as indicated in the legend. In (a) and (b), $n_d$ = 1.8 and $\lambda$ = 1830 nm. The shaded region indicates angles below $\theta_c$, here 39 degrees corresponding to the real part of $n_{CdO}$ = 0.63.

# Supplementary Note 5: Note on efficiency and limitations of the design

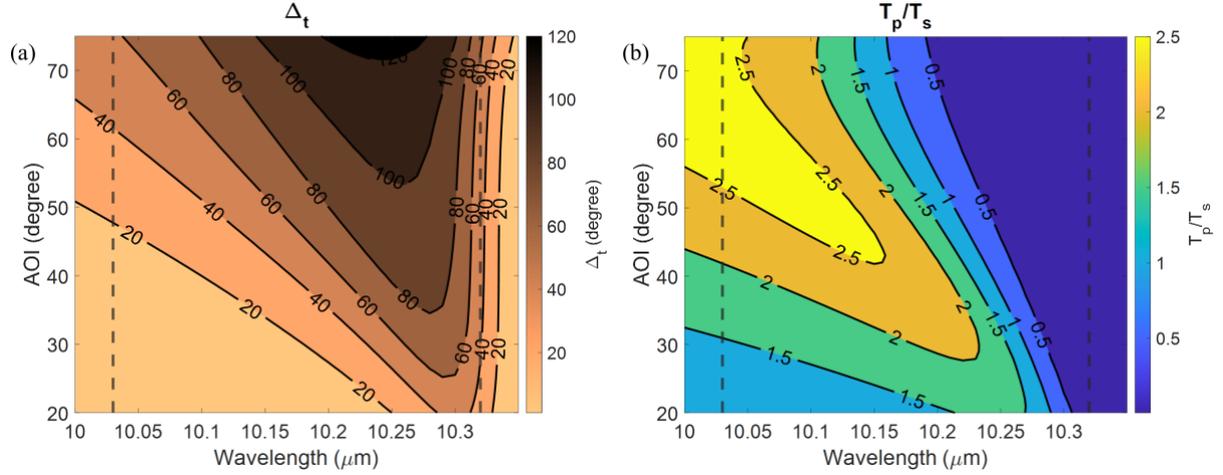

**Figure S5**: Variation of (a) $\Delta_t$ and (b) $T_p/T_s$ (ratio of transmittances) in the dielectric-coated OT configuration (see Figure 4 of main text) as a function of wavelength and AOI.

Earlier works have shown that the efficiency of linear-to-circular polarization conversion under OT is highest when *s* and *p* polarizations tunnel equally, or $T_p = T_s$, where $T$ denotes transmittance [2]. To realize this in OT configuration (the system considered in Figure 4 of the main text), it is necessary to search for parameters such that two conditions are satisfied simultaneously

$$\Delta_t = 90° \quad (S1a)$$
$$T_p = T_s \quad (S1b)$$

The numerical optimization is carried out using a gradient-based constrained optimization routine in MATLAB [3]. Rather than identifying all possible combinations of the relevant parameters that could satisfy eqn. S1, we demonstrate here one possible solution (**Figure S5**). The optimization scheme searches for conditions satisfying eqn. S1 in the vicinity of the ENZ wavelength range of SiC and for AOI above the critical angle, and returns the dielectric thickness $t_d$ and the ENZ thickness $t_{ENZ}$ as results of the optimization. **Figure S5**a,b shows the variation in $\Delta_t$ and $T_p/T_s$ with AOI and wavelength for the optimized values of $t_d$ = 1256 nm and $t_{ENZ}$ = 2732 nm. For these values, $\Delta_t = 89.5°$ and $T_p/T_s$ = 0.984, with the transmitted intensity in each polarization $T_p = T_s \approx 20\%$. Roughly 50% of *p*-polarized light and 15% of *s*-polarized light is absorbed in the device, and the rest is reflected. In other words, 20% of incident linear polarized light is transmitted as circularly polarized light. We further note that losses increase with ENZ layer thickness; however as seen in Figure 3 in the main text, $t_{ENZ} > 3.5\ \mu m$ is required to reach $\Delta_t$ values at or above 90°. To maintain large $\Delta_r$ while reducing ENZ thickness, dielectric coatings of higher refractive index may be used, for example, silicon ($n \approx 3.4$) or germanium ($n \approx 4$). Finally, we note that there is always a limitation to the broadband operation (for e.g., using ultrashort pulses) due to the strong dispersion of the ENZ medium. This is evident from the variation of both $\Delta_t$ and $T_p/T_s$ with wavelength in **Figure S5**. While it is possible to identify conditions where the variation of $\Delta_t$ with wavelength is zero (i.e., a zero value of slope w.r.t the wavelength), the variation in $T_p/T_s$ does not show such behaviour. This would mean even if a relative phase shift of $\Delta_t = 90º$ can be achieved over a range of wavelengths, the variation in $T_p/T_s$ would give rise to elliptical polarization that varies with the wavelength.